\documentclass[12pt]{article}
 
\def\Journal#1#2#3#4{{#1} {\bf #2}, #3 (#4)}

\def\PLB{{\em Phys. Lett.}  B}

\def\PRD{{\em Phys. Rev.} D}
\def\ZPC{{\em Z. Phys.} C}

\def\CPC{{\em Computer Phys. Commun.}}
 
{\end{list}}
\newcounter{enumct}

\newlength{\abstwidth}
\begin{document}
 
\sloppy
 
\pagestyle{empty}
 
\begin{flushright}
LU TP 99--26\\
RAL-TR-1999-065\\
hep-ph/9909346\\
September 1999
\end{flushright}
 
\vspace{\fill}
\begin{center}
{\LARGE\bf P}{\Large\bf YTHIA~}{\LARGE\bf and HERWIG}\\[3mm]
{\LARGE\bf for Linear Collider Physics%
\footnote{To appear in the Proceedings of the International 
Workshop on Linear Colliders, Sitges (Barcelona), Spain, 
April 28 -- May 5, 1999}}\\[10mm]
{\Large Torbj\"orn Sj\"ostrand\footnote{torbjorn@thep.lu.se}}\\[3mm]
{\it Department of Theoretical Physics,}\\[1mm]
{\it Lund University, Lund, Sweden}\\[5mm]
{\large and} \\[5mm]
{\Large Michael H. Seymour\footnote{M.Seymour@rl.ac.uk}}\\[3mm]
{\it Rutherford Appleton Laboratory, Chilton,}\\[1mm]
{\it Didcot, Oxfordshire, OX11 0QX, U.K.}\\[1mm]
\end{center}
 
\vspace{\fill}
 
\begin{center}
{\bf Abstract}\\[2ex]
\begin{minipage}{\abstwidth}
An overview is given of general-purpose event generators,
especially {\sc Pythia} and HERWIG. The current status is 
summarized, some recent physics improvements are described, 
and planned future projects are outlined. 
\end{minipage}
\end{center}
 
\vspace{\fill}
 
\clearpage
\pagestyle{plain}
\setcounter{page}{1}

In order to produce events that can be used for Linear Collider 
physics and detector studies, the structure of the basic 
generation process is \\
1) selection of the hard subprocess kinematics, \\
2) resonance decays that (more or less) form part of the hard 
subprocess (such as $W$, $Z$, $t$ or $h$), \\
3) evolution of QCD parton showers (or, alternatively, the
use of higher-order matrix elements), \\
4) hadronization, and \\ 
5) normal decays (of hadrons and $\tau$'s mainly). \\
Additional aspects, of interest for linear colliders, include \\
6) beamstrahlung (often handled by an interface to 
CIRCE \cite{circe}), \\
7) initial-state QED radiation, e.g. formulated in shower 
language, \\
8) the hadronic behaviour of photons (involving topics such
as the subdivision into direct and resolved photons, VMD and
anomalous ones, parton  distributions of real and virtual 
photons, initial-state QCD radiation, beam remnants of resolved 
photons and even the possibility of multiple interactions in 
those remnants), and \\
9) QCD interconnection effects, e.g. modeled by colour 
rearrangement and Bose--Einstein \cite{intercon}. \\
Finally, since a chain is never stronger than its weakest link, 
one must add\\
10) the forgotten or unexpected. 
  
The historical reason for developing general-purpose generators
has often been an interest in QCD physics: initial- and final-state
cascades, hadronization, underlying events, and so on. However,
once these tools have been developed for simple processes such as
$\gamma^*/Z^0$ production, their generalization to other processes
appears a natural task. There exists three commonly 
used general-purpose generators: {\sc Pythia} \cite{pythia},
HERWIG \cite{herwig} and ISAJET \cite{isajet}. Their main limitation
is that normally only leading-order processes are included,
with higher-order QCD and QED corrections included by showers,
but no weak corrections at all. Furthermore, the nonperturbative
QCD sector is not solved, so hadronization aspects are based on 
models rather than on theory.

Over the years, a long list of physics processes have been added 
to the programs. These cover topics such as hard and soft QCD,
heavy flavours, DIS and $\gamma\gamma$, electroweak production of 
$\gamma^*/Z^0$ and $W^{\pm}$ (singly or in pairs), production of a 
light or a heavy Standard 
Model Higgs, or of various Higgs states e.g. in Supersymmetric (SUSY) 
models, SUSY particle production (sfermions, gauginos, etc.), 
technicolor, new gauge bosons, compositeness, leptoquarks, and so on. 
The most basic processes are included in all the generators, while
the selection diverges for exotic physics. Even when a process formally 
is the same, generators may be based on different theory frameworks 
(e.g. calculation of SUSY parameters) or approximation schemes, and 
are thus not expected to agree completely with each other.
Comparisons between several generators thus are helpful to
assess uncertainties (and, of course, also to find bugs).

The {\sc Pythia} 6.1 program was released in March 1997,
based on a merger of {\sc Jetset} 7.4, {\sc Pythia} 5.7 and   
{\sc SPythia} \cite{spythia}. Main authors are 
T. Sj\"ostrand and \mbox{S.~Mrenna}. New subversions are released
once every few months --- the current one is 6.129, with a size
of about 49\,000 lines of code. The code itself, including manuals
and sample main programs, can be found on\\ 
{\tt http://www.thep.lu.se/$\sim$torbjorn/Pythia.html}.\\
Relative to previous versions, the main news in {\sc Pythia} 6.1
are the transition to double precision throughout and the
new treatment of supersymmetric
processes and particles. Also many other processes have been
added, e.g. for Higgs and technicolor. Colour rearrangement
options for $W^+W^-$ are now included in the code, and the
Bose--Einstein routine has been expanded with many new options.
A new machinery is being built up for real
and virtual photon fluxes and cross sections \cite{christer}. 
An alternative description of popcorn baryon production is 
available \cite{patrik}. New standard interfaces are available 
that should ease the task of matching to external generators of
two, four and six fermions. Among other points, of less relevance 
for $e^+e^-$, one may note the addition of QED radiation off an 
incoming muon, newer parton distributions, and an energy-dependent 
$p_{\perp\mathrm{min}}$ for multiple interactions.

The current HERWIG 5.9 is from July 1996, and has a size of
about 21\,400 lines. Authors are G. Marchesini, B.R. Webber, 
G. Abbiendi, I.G. Knowles, M.H. Seymour and L. Stanco.
Code, manuals and related programs may be found on\\ 
{\tt http://hepwww.rl.ac.uk/theory/seymour/herwig/}.\\
The new version 6.1 is just about to be released, with
G. Corcella, S. Moretti, K. Odagiri and P. Richardson added to the
list of collaborators. The main new improvement is the introduction
of supersymmetric processes within a general MSSM framework,
so far only for hadron collisions, however. Mass and decay spectra 
are not generated intrinsically; instead they are read from a data 
file, e.g. generated by ISAJET/ISASUSY. All $R$-parity conserving
$2 \to 2$ sparticle production subprocesses are available and,
unlike {\sc Pythia} and ISAJET, also all resonant $R$-parity
violating $2 \to 2$ subprocesses and decays. Sparton showering
is not yet included. Most resonances decay isotropically, i.e.
spin correlations are not systematically included. Among other news
one may note a comprehensively expanded set of $2 \to 1$, $2 \to 2$  
and $2 \to 3$ Higgs production subprocesses. An $e^+e^- \to 4$ jets
matrix-element option has been added, the JIMMY generator
for multiparton scattering has been incorporated and improved,
the treatment of $\gamma^*\gamma^*$ and $\gamma$ remnants improved,
and beamstrahlung included by an interface to CIRCE. 

Generator progress is in many directions, and the growth is largely
organic. One main theme in recent times, that will continue to be of
importance, is the gradual improvement of the matching between
higher-order matrix-element information and the parton-shower
language. This is required to obtain an accurate description
of event properties, since each approach has its advantages and 
disadvantages: the former is favoured for the emission of a few
widely separated partons, while the latter is likely to do 
better for multiple emissions at small separations.

One example is the improvement of the description of initial-state
photon radiation in single-$\gamma^*/Z^0$ production in {\sc Pythia}, 
which is a by-product of the study of
$W^{\pm}/\gamma^*/Z^0$ production in hadron colliders \cite{gabriela}. 
The basic idea is to map the kinematics between the parton-shower and 
matrix-element descriptions, and to find a correction factor that
can be applied to hard emissions in the shower so as to bring
agreement with the matrix-element expression. Some simple algebra
shows that, with the {\sc Pythia} shower kinematics definitions,
the two emission rates disagree by a factor
\[
R_{ee \to \gamma Z}(\hat{s},\hat{t}) = 
\frac{(\mathrm{d}\hat{\sigma}/\mathrm{d}\hat{t})_{\mathrm{ME}} }%
     {(\mathrm{d}\hat{\sigma}/\mathrm{d}\hat{t})_{\mathrm{PS}} } = 
\frac{\hat{t}^2+\hat{u}^2+2 m_Z^2\hat{s}}{\hat{s}^2+m_Z^4} ~, 
\]
where $\hat{s}$, $\hat{t}$ and $\hat{u}$ are the standard
Mandelstam variables and $m_Z$ represents the (actual) mass of the 
$s$-channel resonance. This factor is always between $1/2$ and 1. 
The shower can therefore be improved in two ways, relative to the 
old description. Firstly, the maximum virtuality of emissions is 
raised from $Q^2_{\mathrm{max}} \approx m_Z^2$ to 
$Q^2_{\mathrm{max}} = s$, i.e. the shower is allowed to populate the 
full phase space. Secondly, the emission rate for the first (which 
normally also is the hardest) emission on each side is corrected by 
the factor $R(\hat{s},\hat{t})$ above, so as to bring agreement with 
the matrix-element rate in the hard-emission region.

Another example of a shower improvement is the description of gluon
radiation in top decay in HERWIG \cite{gennaro}. The showering of top
decay is done in the top rest frame, where the $W$ and $b$ are
going out back-to-back. In this frame, the gluon emission off the $b$ 
should be smoothly suppressed at large angles relative to the $b$ 
direction, but in HERWIG this is approximated by a sharp step at
90$^{\circ}$. Thus the $W$ hemisphere is left completely empty of 
gluons, while the $b$ one is fully populated. In this kind of 
``dead come approximation'', the total amount of radiation is 
about right, but the angular distribution can be badly wrong.
The HERWIG improvement consists of two parts. A {\em hard} 
correction is applied in the ``dead'' region, where tree-level 
matrix elements are used to populate it (corresponding to roughly 
3\% of the decays). A {\em soft} correction is applied to the populated
region, by a reweighting of emissions, to ensure that the 
kinematical distribution of the hardest emission in the parton shower 
agrees with the tree-level matrix elements \cite{mike}. These
corrections can be very important, especially close to 
threshold \cite{lynn}. Matrix-element corrections to top production 
can also be important, and work is here in progress.

Finally, a word about the future. Both {\sc Pythia} and HERWIG
continue to be developed and supported. On the physics side,
there is a continuous need to increase and improve the support 
given to different physics scenarios, new and old, and many areas 
of the general QCD machinery for parton showers and hadronization 
may require further improvements. On the technical side, the main
challenge is a transition from Fortran to more modern computer 
languages, in practice meaning C++. There are several arguments for 
such a transition. One is that the major labs, such as SLAC, 
Fermilab and CERN have decided to discontinue Fortran support
and go over to C++ as main language. Another is that C++ offers
an educational and professional continuity for students: they
may know it before they begin physics, and they can use it after 
they quit. For experts, C++ is a better programming language.
For the rest of us, user-friendly interfaces should still make life 
easier.

Studies have now begun. The {\sc Pythia} 7 project was formally 
started in January 1998, with L. L\"onnblad as main responsible.
What exists today is a strategy document \cite{leif}, and code for the 
event record and the particle object. The particle data and other 
data base handling is in progress, as is the event generation handler 
structure. The first piece of physics, the string fragmentation scheme,
is being implemented by M. Bertini. The hope its to have a 
``proof of concept'' version soon, and much of the current 
{\sc Pythia} functionality up and running by the end of 2000.
It will, however, take some further time after that to provide a 
program that is both more and better than the current {\sc Pythia} 
version. HERWIG is currently lagging behind, but a plan has been
formulated for a C++ version that would simultaneously offer a 
significantly improved physics content. Recently the PPARC in the
U.K. approved an application for two postdoc-level positions devoted 
full-time to this project, which therefore will start soon.

A copy of the transparencies of this talk, including all the figures
not shown here (for space reasons) may be found on\\ 
{\tt http://www.thep.lu.se/$\sim$torbjorn/talks/sitges99mc.ps}.\\
The talk by F. Paige contains complementary information
on SUSY simulation \cite{paige}.

\end{document}